\documentstyle[osa,12pt]{revtex}

\input{tcilatex}

\begin{document}

SYMMETRY BACKGROUND OF THE STABILITY OF THE

SECOND 0$^{+}$ STATE OF $^4$He

\hspace{1.0in}

C.G.Bao

Center of Theoretical Nuclear Physics, National Laboratory of Heavy Ion
Accelerator, Lanzhou, 730000, China

and

Department of Physics, Zhongshan University, Guangzhou, 510275, China

\hspace{1.0in}

ABSTRACT: It was found in this paper that the dominant component of the wave
function of the second 0$^{+}$ state of $^4$He and the corresponding
component of the $^3$H+p channel have different spatial permutation
symmetries and different parities. This fact will hinder the wave function
to extend from the interior region to the outgoing channel and will results
in having a narrower width.

\hspace{1.0in}



\hspace{1.0in}

\hspace{1.0in}

The few-nucleon systems usually have very few bound states but they might
have a few resonances $^{[1,2]}$. The investigation of the resonances is an
important way to understand the underlying physics of these systems. An
important feature of a resonance is its width. Some of them are very narrow,
some are broad. Usually, the broadness is attributed to the dynamics. Of
course this is true, because if the dynamical parameters vary, the resultant
widths would vary in accord. However, in addition to the dynamics, the
symmetry might be also important. In this paper, it is demonstrated how the
stability of an excited state and thereby its width is affected by symmetry.
Specifically, four-nucleon systems are involved in the discussion.

Based on an R-matrix analysis, a number of resonances of $^4He$ have been
suggested . The existence of the second 0$^{+}$ state as a resonance with a
width about 390 keV is well recognized. However, the other suggested
resonances have not yet been widely accepted. Evidently, if the width of a
resonance is very broad, the discrimination of the resonance would be very
difficult or even impossible. In what follows we shall discuss this problem
in coordinate space.

It is noted that in a region where all the norms of the Jacobi vectors are
smaller than a given reasonable value, then this region is called the
interior region; whereas in a region where the norm of one and only one
Jacobi vector is unlimited, then the region is called a 2-body channel, and
so on. On the other hand, the wave function of a state is in general
composed of different components (as we shall see). If all the components
are confined in the interior region, then the state is obvious a bound
state. However, if some components extend outward along a channel, then it
is a resonance. When the major components are confined in the interior
region while some minor components leak into the channel, then the state is
nearly bound and the width is narrow. Alternatively, if the major components
extend along a channel without being hindered, then the state is far from
stable and the width is broad. In other words, the accessibility of the
channels to the components is crucial to the stability of a state. The
accessibility is not only determined by energy but also by symmetry. In what
follows we shall discuss how a state is decomposed into components, and
which components are allowed to enter a specific channel.

For $^4$He there are five channels; namely the $^3$H+p channel at 19.815
MeV, the $^3$He+n channel at 20.578 , the d+d channel at 23.848, the d+n+p
channel at 26.072, and the 2n+2p channel at 28.30. Let us first consider the
lowest $^3$H+p channel. It is noted that the permutation symmetry of the
spatial wave function of $^3$H is dominated by $\lambda =\{3\}$ $^{[3]}$.
From the outer product \{3\}$\otimes \{1\}$=\{4\}+\{31\}, we know that the
spatial wave function of the $^3$H+p channel contains mainly the \{4\} and
\{31\} components. On the other hand, since both the spin and isospin of $^3$%
H are equal to 1/2, the total spin S and the total isospin T\ of the channel
can be either 0 or 1. When S (T)=0, the permutation symmetry of the
(iso)spin-state is $\{22\}$ , while when S\ (T)=1, it is \{31\}. The
spin-isospin space is an inner product space of the spin space and isospin
space. From the reduction of the inner product space$^{[4]}$, it is known
that, when [S,T]=[0,0], the \{4], \{22\}, and \{1$^4$\} symmetries are
contained in the spin-isospin space; when [S,T]=[0,1] or [1,0], the \{31\}
and \{211\} are contained; when [S,T]=[1,1], the \{4\}, \{31\}, \{22\}, and
\{211\} are contained. Let $\Psi _{channel}$ denote the wave function of the
channel with a given total angular momentum J, parity $\Pi $, and T. Since,
as a fermion system, the total wave function must be antisymmetric, a
component of the spatial wave function with the symmetry $\lambda $ must be
coupled with a spin-isospin partner with exactly the conjugate symmetry of $%
\lambda $. Thus we have

$\Psi _{channel}=\sum_{L,S}\Psi _{channel}^{L,S}$\hspace{1.0in}(1)

\hspace{1.0in}

$\Psi _{channel}^{L,S}=\sum_i(F_{LS\Pi }^{\{4\},i}\cdot \chi
_{S,T}^{\{1^4\},i})_J$ +$\Psi _\varepsilon $\qquad (if [S,T]=[0,0])%
\hspace{1.0in}(2.1)

\hspace{1.0in}

$\Psi _{channel}^{L,S}=\sum_i(F_{LS\Pi }^{\{31\},i}\cdot \chi
_{S,T}^{\{211\},i})_J$ +$\Psi _\varepsilon $\qquad (if [S,T]=[0,1], [1,0],
or [1,1])\hspace{1.0in}(2.2)

\hspace{1.0in}

Where L is the total orbital angular momentum, L and S are coupled to J, $%
F_{LS\Pi }^{\lambda ,i}$ is a function of the spatial coordinates with the
symmetry $\lambda $, i specifies a basis function of the $\lambda -$%
representation, $\chi _{S,T}^{\stackrel{\sim }{\lambda },i}$ is a
spin-isospin state with the symmetry $\stackrel{\sim }{\lambda }$, $\lambda $
and $\stackrel{\sim }{\lambda }$ are conjugate with each other. $\Psi
_\varepsilon $ is a small component arising from the minor component of $^3$%
H with spatial permutation symmetry not equal to \{3\}. In (2) the coupling
of $\lambda $ and $\stackrel{\sim }{\lambda }$ is the only choice to assure
the antisymmetrization (if $\lambda $ is not coupled with $\stackrel{\sim }{%
\lambda }$ , the related terms will be automatically cancelled during the
antisymmetrization).

Let $l$ be the partial wave of the relative motion between the $^3$H and the
p. Since $^3$H has mainly a zero orbital angular momentum and an even parity 
$^{[3]}$, we have $\Pi =(-1)^l=(-1)^L$. This is an additional condition
imposed on (2).

On the other hand, in the interior region a wave function with a given J, $%
\Pi $, and T can be in general decomposed into components as

$\Psi _{in}=\sum_{LS,i}(G_{LS\Pi }^{\lambda ,i}\cdot \chi _{S,T}^{\stackrel{%
\sim }{\lambda },i})_J$\hspace{1.0in}(3)

Where $G_{LS\Pi }^{\lambda ,i}$ is a function of the spatial coordinates
with the $\lambda $ symmetry. It was found in the ref. [5,6] that there are
two kinds of nodal surfaces existing in the spatial wave functions, namely
the inherent nodal surfaces (INS) and the dynamical nodal surfaces (DNS).
The first kind is dynamics-independent (they can not be moved by adjusting
dynamical parameters), they arise simply from the constraints of symmetry.
The second kind does not have symmetry background but is strongly
dynamics-dependent (they move in accord with the variation of dynamical
parameters), they arise from the requirement of orthogonality. For a series
of mutually orthogonal spatial wave functions having the same L, $\Pi $, and 
$\lambda $, it was found that all of them have exactly the same INS (if they
have), but the number and locations of the DNS\ are different in different
wave functions to assure the orthogonality. The higher the energy, the more
the DNS are contained. For the lowest one of this series, the wave function
does not contain the DNS but it might contain the INS. Whether the INS exist
depends on the common L, $\Pi $, and $\lambda $ of the wave functions $^{[6]}
$. Among all the $G_{LS\Pi }^{\lambda ,i}$ components in eq.(3), some of
them contain the INS, but some do not. Evidently, those contain the INS
would not be preferred by low-lying states because a nodal surface would
cause a great increase in energy. In other words, those do not contain the
INS\ would be the dominant components. For 4-nucleon systems, which $%
G_{LS\Pi }^{\lambda ,i}$ contains the INS and which does not has been
clarified in [6]. Thereby it was deduced that the second 0$^{+}$state is
dominated by the $G_{LS\Pi }^{\lambda ,i}$ component with L=1, S=1, and $%
\lambda =\{211\}$. Thus, for the second 0$^{+}$ state, eq.(3) can be
rewritten as

$\Psi _{in}=\sum_i(G_{11+}^{\{211\},i}\cdot \chi _{1,0}^{\{31\},i})_J+\Psi
_\varepsilon $\hspace{1.0in}(4)

where the $\Psi _\varepsilon $ is small as before. On the other hand, from
eq.(2.2) and from the requirement $\Pi =(-1)^L$, for the case of T=0, S=1,
and L=1, we have

$\Psi _{channel}^{L,S}=$ $\sum_i(F_{11-}^{\{31\},i}\cdot \chi
_{1,0}^{\{211\},i})_J+\Psi _\varepsilon $\hspace{1.0in}(5)

.Comparing (4) with (5), it is clear that the main component of $\Psi _{in}$
does not match the corresponding $\Psi _{channel}^{L,S}$ both in $\lambda $
and in parity. This might greatly hinder the wave function to extend from
the interior to the channel . It is noted that the second 0$^{+}$ state lies
at 20.21 MeV$^{[1]}$. For this state the $^3$H+p is the only outgoing
channel available. Since this state is greatly hindered by symmetry to enter
the channel, it should have a narrower width as confirmed in the
literatures. Incidentally, the minor component $\Psi _\varepsilon $ in (4)
would contain the L=0, S=0, and $\lambda =\{4\}$ component, this component
can entend into the channel without being hindered (cf. eq.(2.1)) and
therefore would contribute to the width.

Let us now study the stability of the other excited states of $^4$He lower
or equal to 28.31 MeV (very close to the 2n+2p threshold). It was found in
[6] that all these states are dominated by L=1. The other (approximately)
good quantum numbers found in [6] together with a serial number k are given
in Table 1. For the high-lying k=9 and 10 states, they are also hindered to
enter the $^3$H+p channel because they have an even-parity and $\lambda
=\{211\}$ as well. On the other hand, the k=2 to 8 states have an odd-parity
and $\lambda =\{31\}$. Thus their main component of $\Psi _{in}$ match the
corresponding $\Psi _{channel}^{L,S}$, therefore their wave functions can
gently extend into the channel without being hindered. Accordingly, their
widths are broader.

The accessibility of the $^3$He+n channel is the same as the $^3$H+p
channel, but the threshold of the former is a little higher.

For the d+d channel, we have T=0; S=0,1, and 2; and the spatial permutation
symmetry is $\{2\}\otimes \{2\}=\{4\}+\{31\}+\{22\}$. Thus the wave function
of the channel can be written as

$\Psi _{channel}^{L,S}=\sum_i[(F_{LS\Pi }^{\{4\},i}\cdot \chi
_{S,T}^{\{1^4\},i})_J$ +$(F_{LS\Pi }^{\{22\},i}\cdot \chi
_{S,T}^{\{22\},i})_J]+\Psi _\varepsilon $\hspace{1.0in}(if S=0)\hspace{1.0in}%
(6.1)

\hspace{1.0in}

$\Psi _{channel}^{L,S}=\sum_i(F_{LS\Pi }^{\{31\},i}\cdot \chi
_{S,T}^{\{211\},i})_J+\Psi _\varepsilon $ \hspace{1.0in}(if S=1)%
\hspace{1.0in}(6.2)

\hspace{1.0in}

$\Psi _{channel}^{L,S}=\sum_i(F_{LS\Pi }^{\{22\},i}\cdot \chi
_{S,T}^{\{22\},i})_J+\Psi _\varepsilon $\hspace{1.0in}(if S=2)\hspace{1.0in}%
(6.3)

\hspace{1.0in}

In eq.(6) additional conditions T=0 and $\Pi =(-1)^L$ are required. It is
noted that all the $\Psi _{in}$ of the T=0 excited states are dominated by
S=1 and L=1 component (cf. Table 1). From (6.2) the corresponding channel
wave function reads

$\Psi _{channel}^{1,1}=$ $\sum_i(F_{11-}^{\{31\},i}\cdot \chi
_{1,0}^{\{211\},i})_J+\Psi _\varepsilon $\hspace{1.0in}(7)

Since the $\Psi _{in}$ of the k=9 and 10 states have an even-parity and
mainly have $\lambda =\{211\},$ therefore they are hindered to enter the
channel. Whereas the k=6 state is not hindered because it has an odd-parity
and mainly has $\lambda =\{31\}$.

The accessibility of the channels to the ten excited states are summarized
in Table 1. It is clear from this table that , except the k=1 state (the
second 0$^{+}$ state), all the other states can enter a number of channels
without being hindered. Thus the widths of these states should be
considerably broader than the k=1 state, therefore they are difficult (or
even impossible) to be discriminated. Nonetheless, the k=9 state has only
one channel to escape without being hindered. Thus the width of this state
should be relatively narrower than the other eight.. This 2$^{+}$ state with
T=0 might be easier to be discriminated, this is an open problem.

\hspace{1.0in}

\begin{tabular}{|c|c|c|c|c|c|c|c|c|c|}
\hline
k & J$^\Pi $ & S & T & $\lambda $ & $^3$H+p & $^3$He+n & d+d & d+n+p & 
2n+2p \\ \hline
1 & 0$^{+}$ & 1 & 0 & \{211\} & H & P & P & P & P \\ \hline
2 & 0$^{-}$ & 1 & 0 & \{31\} &  &  & P & P & P \\ \hline
3 & 2$^{-}$ & 1 & 0 & \{31\} &  &  & P & P & P \\ \hline
4 & 2$^{-}$ & 1 & 1 & \{31\} &  &  & P & P & P \\ \hline
5 & 1$^{-}$ & 0+1 & 1 & \{31\} &  &  & P & P & P \\ \hline
6 & 1$^{-}$ & 1 & 0 & \{31\} &  &  &  & P & P \\ \hline
7 & 0$^{-}$ & 1 & 1 & \{31\} &  &  & P & P & P \\ \hline
8 & 1$^{-}$ & 0+1 & 1 & \{31\} &  &  & P & P & P \\ \hline
9 & 2$^{+}$ & 1 & 0 & \{211\} & H & H & H &  & P \\ \hline
10 & 1$^{+}$ & 1 & 0 & \{211\} & H & H & H &  &  \\ \hline
\end{tabular}

Table 1, The accessibility of the channels to the excited states of $^4$He.
The levels are ordered in increasing energy with a serial number k. The
approximately good quantum numbers are found in ref.[6], where $\lambda $ is
a representation of the $S_4$ group for the spatial wave functions. The
''P'' implies that the associated state is prohibited by energy or isospin
conservation to enter the channel. The ''H'' implies that the associated
state is hindered by symmetry to enter the channel (i.e., the main component
of $\Psi _{in}$ does not match the corresponding $\Psi _{channel}^{L,S}$ in $%
\lambda $ and(or) in parity). An empty block implies that the associated
state can enter the channel without being hindered.

\hspace{1.0in}

In summary, the symmetry background of the stability of the excited states
of $^4$He has been studied in this paper. The procedure of analysis is quite
general and can be generalized to study other few-body systems $^{[7]}$.

\hspace{1.0in}

ACKNOWLEDGMENT: This work is supported by the NSFC and by a fund from the
National Educational Ministry.

\hspace{1.0in}

REFERENCES

1, F.Ajzenberg-Selove, Nucl. Phys. A490 (1988) 1

2, A. Csoto and G.M. Hale, Nucl. Phys. A631 (1998) 783c

3, Y.Wu, S.Ishikawa, and T.Sasakawa; Few-Body Systems, 15 (1993) 145

4, C. Itzykson and M. Nauenberg, Rev. Mod. Phys. 38 (1966) 95

5, C.G.Bao, Few-Body Systems, 13 (1992) 41

6, C.G.Bao, Nucl. Phys. A. 637 (1998) 520

7, C.G.Bao and Y.X.Liu, Phys. Rev. Lett. 82 (1999) 61

\end{document}